\documentclass[aps,prd,twocolumn,preprintnumbers,nofootinbib,showpacs]{revtex4}
\usepackage{graphicx}
\usepackage{epsfig}
\usepackage{amssymb}
\usepackage{color}
\usepackage{epstopdf}
\usepackage{dcolumn}
\usepackage{amsmath}
\usepackage{bm}
\usepackage{dcolumn}
\usepackage{txfonts}
\begin{document}

\title{Quark masses and strong coupling constant in 2+1 flavor QCD}

\author{
Y. Maezawa$^{\rm a}$ and
P. Petreczky$^{\rm b}$, 
}
\affiliation{
$^{\rm a}$ Yukawa Institute for Theoretical Physics, Kyoto University, Kyoto 606-8317, Japan\\
$^{\rm b}$ Physics Department, Brookhaven National Laboratory, Upton, New York 11973, USA 
}

\begin{abstract}
We present a determination of the strange, charm and bottom quark masses as well as 
the strong coupling constant in 2+1 flavor lattice QCD simulations using highly improved staggered quark action.
The ratios of the charm quark mass to the strange quark mass and the bottom quark mass 
to the charm quark mass are obtained from the meson masses calculated on the lattice and found to be $m_c/m_s=11.877(91)$ and $m_b/m_c=4.528(57)$ in the continuum limit.
We also determine the strong coupling constant and the charm quark mass 
using the moments of pseudoscalar charmonium correlators: $\alpha_s(\mu=m_c)=0.3697(85)$ and 
$m_c(\mu=m_c)=1.267(12)$ GeV. Our result for $\alpha_s$ corresponds to the determination of the strong coupling constant
at the lowest energy scale so far and is translated to the value $\alpha_s(\mu=M_Z,n_f=5)=0.11622(84)$.
\end{abstract}
\date{\today}

\maketitle

\section{Introduction}
Accurate determination of QCD parameters received a lot of attention in recent years.
Lattice QCD calculations play an important role in this quest. The precise knowledge
of the QCD parameters is important for testing the prediction of the standard model.
One prominent example is the sensitivity of Higgs branching ratios to the heavy quark masses
and the strong coupling constant \cite{Lepage:2014fla}. While several precise determinations
of the heavy quark masses and the strong coupling constant $\alpha_s$ on the lattice exist,
it is always important to obtain results using different lattice methods to ensure that all
the errors are under control. In the case of $\alpha_s$ different lattice and nonlattice methods
often give quite different results, possibly suggesting that not all the sources of errors are
under control \cite{Moch:2014tta}. In particular, the lattice determinations that use the static
quark antiquark potential lead to smaller
values of $\alpha_s$ \cite{Bazavov:2012ka,Bazavov:2014soa}.
As a result of this the error on the $\alpha_s$ quoted in the most recent Particle Data Group (PDG)
Review update has increased for the first time in many years: $\alpha_s(M_Z)=0.1181(16)$ \cite{PDG15}.
This should be compared to the 2013 PDG value, $\alpha_s(M_Z)=0.1185(6)$.
Lattice QCD offers the possibility to determine the strong coupling constant at relatively
low energy scales. So far the only nonlattice method that offers a low energy determination
of $\alpha_s$ is the analysis of the $\tau$ decay but there are large systematic
uncertainties due to different ways of organizing the perturbative expansion
in this method (see Ref.~\cite{Pich:2016bdg} for a recent work on this topic and
references therein). For certain applications it is important to have the running of the coupling constant
at low energy scales. One example is the comparison of weak coupling and lattice results
in QCD thermodynamics, where the typical scale $\simeq \pi T$ could be as low as $1$ GeV 
\cite{Bazavov:2016uvm,Berwein:2015ayt,Ding:2015fca,Bazavov:2013uja}.

There are also sizable differences in the value of the charm quark masses. The recent determination
of $m_c$ by the HPQCD Collaboration \cite{Chakraborty:2014aca} 
is significantly lower than the value obtained by the ETMC Collaboration
\cite{Carrasco:2014cwa}. Some lattice QCD calculations use 2 or 3 flavors of dynamical quarks
\cite{Durr:2011ed,Davies:2009ih,Yang:2014sea}, 
while others use 4 dynamical flavors \cite{Chakraborty:2014aca,Carrasco:2014cwa,Bazavov:2014wgs}.
Therefore, understanding of the flavor dependence of the charm quark mass is also important. 

Furthermore, nonperturbative determination of the bottom quark mass is a problematic matter in 
the lattice simulations due to the discretization errors caused by powers of $m_h a$, where $m_h$ is the bare mass of the heavy quarks.
However, owing to improvements of discretization of the action as well as simulations with 
smaller lattice spacing using powerful computing resources, it has recently become possible to perform calculations 
with quark masses larger than the charm quark mass.  
The region around the bottom quark mass can be accessed using extrapolations \cite{McNeile:2010ji}.
Several determinations of the quark mass ratio of the bottom to charm have been reported, and a slight 
inconsistency has been found: The ratio recently obtained by the ETMC Collaboration \cite{Bussone:2016iua} 
shows smaller value than that previously determined by the HPQCD Collaborations \cite{McNeile:2010ji,Chakraborty:2014aca}.
Thus the determinations of the bottom quark mass with different setups and approaches are also 
important to provide precise theoretical predictions.

In this paper we report on the calculation of the quark masses and the strong coupling constant 
in 2+1 flavor QCD using highly improved staggered
quark (HISQ) action. 
More precisely, we determine the ratio of the charm quark mass to the strange quark mass and the bottom quark mass 
to the charm quark mass from the pseudoscalar and vector meson masses calculated on the lattice and combined with the experimental inputs.
Furthermore, the strong coupling constant $\alpha_s$ and the charm quark mass 
$m_c(m_c)$ in $\overline{\rm MS}$ renormalization scheme are determined from the moments 
of the pseudoscalar charmonium correlators and the comparison to the corresponding perturbative result.
By using the quark mass ratios together with $\alpha_s$ and $m_c(m_c)$, we also determine the strange and bottom quark masses.

This paper is organized as follows: In Sec.~\ref{sec:lat_setup} we introduce the details of the lattice setup 
and explain our approach to determine the quark mass ratios, the strong coupling constant and the quark masses.
Our main numerical results are discussed in Sec.~\ref{sec:analysis}, including the determination of the physical values of
the charm quark mass and the ratios of the quark masses, as well as the moments of the pseudoscalar charmonium correlators.
In Sec.~\ref{sec:comp} we compare our results for the strong coupling constant and the quark masses with other lattice results.
The paper is concluded in Sec.~\ref{sec:conclusion}.

\section{Lattice setup and details of analysis}
\label{sec:lat_setup}
To determine the quark masses and the strong coupling constant,
we calculate meson masses as well as the moments of pseudoscalar charmonium correlators in 2+1 flavor lattice QCD.
The gauge configuration used in our study has been generated using tree-level improved
gauge action \cite{Luscher:1984xn} and highly improved staggered quark (HISQ) action \cite{Follana:2006rc}
by the HotQCD Collaboration \cite{Bazavov:2014pvz}. The strange quark mass, $m_s$, was fixed to its
physical value, while for the light (u and d) quark masses the value $m_l=m_s/20$ was used.
The later corresponds to the pion mass of $160$ MeV in the continuum limit. Thus, the values
of the light quark masses are slightly larger than the physical value. This small
difference , however, does not lead to any visible effects in the physical observables at
zero temperature, which agree well with the experimental values \cite{Bazavov:2014pvz,Bazavov:2014cta}.
For the valence charm and bottom quarks we use the HISQ action with the so-called
$\epsilon$-term \cite{Follana:2006rc}, which removes the tree-level discretization effects
due to the large quark mass up to ${\cal O}((a m)^4)$. The HISQ action with $\epsilon$-term
turned out to be very effective for treating the charm quark on the lattice
\cite{Follana:2006rc,Bazavov:2014wgs,Bazavov:2014cta,Mohler:2014ksa}.
The lattice spacing in our calculations has been fixed using the $r_1$ scale
defined in terms of the energy of static quark antiquark pair $V(r)$ as
\begin{equation}
\left. r^2 \frac{d V}{d r} \right|_{r=r_1}=1.0.
\end{equation}
We use the value of $r_1$ determined in Ref.~\cite{Bazavov:2010hj}
using the pion decay constant as an input:
\begin{equation}
r_1=0.3106 \, (18) \ {\rm fm}.
\end{equation}
In the above equation all the sources of errors in Ref.~\cite{Bazavov:2010hj} have been
added in quadrature. 
The above value of $r_1$ corresponds to the value of the scale parameter 
determined from the Wilson flow $w_0=0.1749(14)$ \cite{Bazavov:2014pvz}. 
This agrees very well with determination of the Wilson flow parameter by the BMW Collaboration, $w_0=0.1755(18)(4)$ \cite{Borsanyi:2012zs}.
It is also consistent with the HPQCD value $r_1=0.3133(23)(3)$ within errors \cite{Davies:2009tsa}. 
This gives us confidence in our scale setting.

All of the quantities in this paper can be calculated from the meson correlation functions.
In this study, we focus only on the local meson operators which have the same structures in the taste and spin generators of Dirac gamma matrices $\Gamma$.
In particular, we calculate the meson propagators consisting of the pseudoscalar $\Gamma=\gamma_5$ and vector $\gamma_i$ operators.
To obtain the moments of the charmonium correlators (explained in detail below), we calculate the pseudoscalar meson correlators with the point sources.
On the other hand, to determine the bare charm quark mass and quark mass ratios, $m_c/m_s$ and $m_b/m_c$, we utilize the meson correlators obtained with the corner-wall sources, where on a given time slice we set the sources to one at the origin of each $2^3$ cube and to zero elsewhere.
The corner-wall sources enable reduction of the contribution of higher excited states and thus more accurate determination of the ground state masses.
From the meson propagators, we extract the charmonium and bottomonium masses using two type of fits.
The first type of fits includes only the ground state contributions, while the second type of fits
includes the ground state contribution and the first excited state contribution \cite{Lepage:2001ym}. 
The second type of fit allows us to use a larger range in the time direction.
We find that the two fits agree
quite well. We also checked the fit range dependence of the extracted masses and found
it to be small. Any dependence on the fit range that is larger than the statistical
error is treated as a systematic error. 
For the determination of the ratio $m_c/m_s$, we need the mass of the unmixed 
pseudoscalar $s \bar s$ meson mass at the physical point and utilize the lattice results from Ref.~\cite{Bazavov:2014pvz}.

Using the $J/\psi$ and $\eta_c$ masses
obtained for several trial values of the lattice bare quark mass $m_{ct}$, we study the charm quark mass dependence
of the spin averaged mass
\begin{equation}
\overline{M}=\frac{1}{4}( 3 M_{J/\psi}+ M_{\eta_c} ).
\end{equation}
Using $\overline M$ has the advantage that effects of hyperfine splitting, which
are sensitive to discretization errors, cancel out in this combination.
We fit the $m_{ct}$ dependence of $\overline M$ using the linear form
\begin{equation}
\overline{M}= d+ b m_{ct},
\label{eq:Mbar}
\end{equation}
which works very well. 
Then the physical value of the bare charm quark mass can be determined as
\begin{equation}
m_{c0} = \frac{1}{b}\left[ \overline{M} - d_0 \left(\frac{r_1}{a}\right)r_1^{-1}\right]
,\label{eq:mc0}
\end{equation}
where we explicitly expressed $d$ in terms of a dimensionless quantity given in the lattice unit: $d_0=ad$.

The mass of the unmixed pseudoscalar meson
is given by $M_{\eta_{s\bar{s}}}^2=B m_{s0}$.
Then, the mass ratio of the charm to strange quarks can be written as
\begin{equation}
\frac{m_{c0}}{m_{s0}}=\frac{B}{M_{\eta_{s\bar{s}}}^2}\frac{\overline{M}-d}{b} \, .
\end{equation}
By using the $r_1$ scale as well as the values of  $B$, $r_1/a$, $d_0$, and $b$, extracted on the lattice,
the above equation can be rewritten as
\begin{equation}
\frac{m_{c0}}{m_{s0}}=\frac{B_0}{b M_{\eta_{s\bar{s}}} r_1} \left(\frac{r_1}{a}\right) \left[ 
\frac{\overline M}{M_{\eta_{s\bar{s}}}}
-\left(\frac{r_1}{a}\right)\frac{d_0}{M_{\eta_{s\bar{s}}} r_1} \right ],
\label{basic}
\end{equation}
where $B_0=a B$. Since the ratio of the quark masses is scheme and scale independent,
$m_c/m_s=m_{c0}/m_{s0}$, 
the above equation is the basis for our extraction of $m_c/m_s$.
Using the experimental input for the meson masses $\overline M$ and 
$M_{\eta_{s\bar{s}}}$ on one hand and the value of the fit parameters $b$ and $d_0$
obtained on the lattice together with the values of $B_0$ and $r_1/a$ from
Ref.~\cite{Bazavov:2014pvz} on the other hand, we can obtain the value of $m_c/m_s$ 
at each lattice spacing. Next, we have to perform the continuum extrapolation
of this ratio to obtain its physical value. In the next section, we discuss the numerical
details of these steps along the discussion of the corresponding error budget.

A similar approach can be applied to the bottom quark mass and the ratio of the bottom to charm quark mass.
With the meson correlation functions at heavy valence quark masses, 
in general $m_h > m_{c0}$; we also fit the quark mass dependence of pseudoscalar masses with the linear form
\begin{equation}
M_{\eta_h} = d_h + b_h m_h \,.
\label{eq:metah}
\end{equation}
With the experimental value of the $\eta_b$ meson mass, the quark mass ratio can be evaluated as
\begin{eqnarray}
\frac{m_b}{m_c} &=& \frac{M_{\eta_b} - d_h}{\overline{M} - d}\frac{b}{b_h} \nonumber \\
&=& \frac{b}{b_h} \frac{r_1M_{\eta_b} - d_{h0} (r_1/a)}{r_1\overline{M} - d_0 (r_1/a)} \, ,
\label{eq:mbmc}
\end{eqnarray}
where $d_{h0}=ad_h$.
Here, we use the pseudoscalar mass instead of the spin averaged mass because the effects of hyperfine splitting are
quite small compared to the overall mass scale.
Even in the state-of-the-art lattice simulations, it is difficult to obtain the $\eta_b$ mass because 
$am_b  \sim 1.0$, and the discretization errors are significant.
To circumvent this problem, we perform calculations for several values of the valence quark masses that are
smaller than the bottom quark mass and extrapolate to the region of the bottom quark mass. 
If the utilized valence quark masses are too small, this procedure could have systematic uncertainties. 
To investigate such uncertainties, we perform the extrapolations from the data points 
at $am_h<1.0$ to $m_{b0}$ with several mass ranges and estimate discrepancy between the results
obtained using different ranges. These discrepancies are treated as systematic errors. 

Once the lattice charm quark masses $m_{c0}$ corresponding to the physical value have been determined,
we calculate the pseudoscalar charmonium correlator with the valence mass of $m_{c0}$ using point sources. 
Then, we consider the moments of the pseudoscalar charmonium correlator,
which are defined as
\begin{equation}
G_n=\sum_t t^n G(t),~G(t)=a^6 \sum_{\mathbf{x}} (a m_{c0})^2 \langle j_5(\mathbf{x},t) j_5(0,0) \rangle,
\end{equation}
Here, $j_5=\bar \psi \gamma_5 \psi $ is the pseudoscalar current.
To take into account the periodicity of the lattice of temporal size $N_t$, the above definition of the moments can
be generalized as follows:
\begin{equation}
G_n=\sum_t t^n (G(t)+G(N_t-t)).
\label{eq:latmom}
\end{equation}
The moments $G_n$ are finite for $n \ge 4$ ($n$ even) since the correlation function 
diverges as $t^{-4}$ for small $t$. Furthermore, the moments $G_n$ do not
need renormalization because
 the explicit factors of the quark mass are included in
their definition \cite{Allison:2008xk}.
They can be calculated in perturbation theory in $\overline{\rm MS}$ scheme
\begin{equation}
G_n=\frac{g_n(\alpha_s(\mu),\mu/m_c)}{a m_c^{n-4}(\mu)}.
\end{equation}
Here, $\mu$ is the $\overline{\rm MS}$ renormalization scale.
The coefficient $g_n(\alpha_s(\mu),\mu/m_c)$ is calculated up to 4-loop, i.e., up to order $\alpha_s^3$
\cite{Sturm:2008eb,Kiyo:2009gb,Maier:2009fz}.
Given the lattice data on $G_n$, one can extract $\alpha_s(\mu)$ and $m_c(\mu)$ from the above
equation. However, as it was pointed out in Ref.~\cite{Allison:2008xk}, it is more practical to
consider the reduced moments
\begin{equation}
R_n =\left\{ \begin{array}{ll}
G_n/G_n^{(0)} & (n=4) \\
\left(G_n/G_n^{(0)}\right)^{1/(n-4)} & (n\ge6) \\
\end{array} \right.
\label{eq:redmom},
\end{equation}
where $G_n^{(0)}$ is the moment calculated from the free correlation function.
The lattice artifacts largely cancel out in these reduced moments.
It is straightforward to write down the perturbative expansion for $R_n$:
\begin{eqnarray}
R_n &=& \left\{ \begin{array}{ll}
r_4 & (n=4) \\
r_n \cdot \left({m_{c0}}/{m_c(\mu)}\right) & (n\ge6)\\
\end{array}\right. , \\
r_n &=& 1 + \sum_{j=1}^3 r_{nj}(\mu,m_c) \alpha_s^j(\mu).
\end{eqnarray}
For the choice of the renormalization scale $\mu=m_c$, the expansion coefficients
are just simple numbers that have been tabulated, for example, in Ref.~\cite{Chakraborty:2014aca}.
This choice of the renormalization scale has the advantage that the expansion coefficients
are never large.

\begin{table}[tb]
 \begin{center}
 \caption{The gauge couplings ($\beta$), lattice sizes ($N_s^3\times N_t$), and the strange quark masses ($am_s$) used in our calculations, as well as corresponding lattice spacing ($a^{-1}$ [GeV]).
 The number of trajectories (traj.) we use to calculate the charmonium correlation function with the corner-wall sources 
 are also summarized, as well as the results for the bare charm quark masses, $m_{c0}$ in GeV, and ratios of charm to strange quark masses $(m_c/m_s)$. 
 The calculations have been done every five trajectories for $N_t=32$ and 48 and six trajectories for $N_t=64$.}
 \label{tab:param_charm}
 {\renewcommand{\arraystretch}{1} \tabcolsep = 1mm
 \newcolumntype{.}{D{.}{.}{2}}
 \begin{tabular}{ccccccc}
 \hline\hline
$\beta$ & $N_s^3\times N_t$ & $am_s$ & $a^{-1}$ [GeV] & traj. & $m_{c0}$ & $m_c/m_s$ \\
 \hline
6.488 & $32^4$         & 0.0620 & 1.42 & 2500 & 1.0899(23) & 12.586(28) \\
6.515 & $32^4$         & 0.0604 & 1.46 & 2500 & 1.0810(23) & 12.518(28) \\
6.664 & $32^4$         & 0.0514 & 1.69 & 2500 & 1.0407(20) & 12.299(25) \\
6.740 & $48^4$         & 0.0476 & 1.81 & 2440 & 1.0215(18) & 12.162(22) \\
6.880 & $48^4$         & 0.0412 & 2.07 & 2465 & 0.9935(21) & 12.023(26) \\
7.030 & $48^4$         & 0.0356 & 2.39 & 1530 & 0.9673(21) & 11.917(27) \\
7.150 & $48^3\times64$ & 0.0320 & 2.67 & 2406 & 0.9471(25) & 11.926(32) \\
7.280 & $48^3\times64$ & 0.0284 & 3.01 & 2376 & 0.9289(25) & 11.886(34) \\
7.373 & $48^3\times64$ & 0.0250 & 3.28 & 1206 & 0.9161(27) & 11.832(35) \\
7.596 & $64^4$         & 0.0202 & 4.00 & 1200 & 0.8878(34) & 11.850(46) \\
7.825 & $64^4$         & 0.0164 & 4.89 & 1200 & 0.8679(57) & 11.930(80) \\
 \hline
 \end{tabular}}
 \end{center}
\end{table}

\section{Numerical analysis and continuum extrapolations}
\label{sec:analysis}
\subsection{Determinations of $m_{c0}$ and ratios $m_c/m_s$ and $m_b/m_c$}
To obtain the value of $m_{c0}$ as well as $m_c/m_s$ and $m_b/m_c$ 
at each lattice spacing, we need physical masses: $\overline{M}$, $M_{\eta_{s\bar{s}}}$, and $M_{\eta_b}$.
We directly utilize the experimental value $M_{\eta_b} = 9.3980(32)$ GeV from PDG \cite{PDG15},
 whereas to specify $\overline M$ we take the values of $\eta_c$ and $J/\psi$ masses
 from PDG and obtain $\overline{M}=3.067$ GeV. 

The gauge configurations used in our study are summarized in Table \ref{tab:param_charm}, together
with the number of analyzed trajectories and the corresponding lattice spacings.
The statistical errors on meson masses and more generally on meson correlation functions
and their moments have been estimated using jackknife analysis. We varied the jackknife 
bin size and checked that the estimated statistical errors do not change significantly. 
Therefore, the analysis does not suffer from the effects of autocorrelations.

Our calculation neglects
disconnected diagrams and electromagnetic effects. The effect of
disconnected diagrams on the ground state charmonium is know to be a few
MeV \cite{Levkova:2010ft}. Electromagnetic effects are of similar size \cite{Chakraborty:2014aca}.
Therefore, following Ref.~\cite{Chakraborty:2014aca}, we assign an error of $3$ MeV to the value
of $\overline M$. 
On the other hand, to estimate the unmixed pseudoscalar $s\bar s$ meson mass, 
we use the leading order chiral perturbation theory, $M_{\eta_{s\bar{s}}}=\sqrt{2 M_K^2 - M_{\pi}^2}$.
We need, however, to take into account the breaking of the isospin symmetry and the electromagnetic
effects, which are neglected in our calculations.
Following Ref.~\cite{Aubin:2004fs} for the pion and kaon masses, we write
\begin{eqnarray}
&
\displaystyle
M_{\pi}^2=M_{\pi^0}^2, \\
&
\displaystyle
M_{K}^2=\frac{1}{2}(M_{K^0}^2+M_{K^+}^2-(1+\Delta_E)(M_{\pi^+}^2-M_{\pi^0}^2) ).
\end{eqnarray}
The parameter $\Delta_E$ characterizes the violation of the Dashen theorem stating
that in the chiral limit the electromagnetic corrections to $M_{K^+}$ and $M_{\pi^+}$
are the same, while there are no electromagnetic corrections to $M_{\pi^0}$ and $M_{K^0}$.
The value of $\Delta_E$ was determined to be $\Delta_E=0.84(25)$ \cite{Bijnens:1996kk}.
There is also a very recent lattice determination of this parameter \cite{Fodor:2016bgu}.
Here, however, we follow Ref.~\cite{Aubin:2004fs} and use a more conservative approach 
varying $\Delta_E$ from $0$ to $2$. We find 
\begin{equation}
M_{\eta_{s\bar{s}}}=686.00(92)~{\rm MeV}, 
\end{equation}
where the central value corresponds to $\Delta_E=0$. This value is in excellent agreement
with the HPQCD determination, $M_{\eta_{s\bar{s}}}=685.8(3.8)(1.2)$ MeV \cite{Davies:2009tsa}, making us
confident that the value based on leading order chiral perturbation theory is 
accurate.

To associate the absolute scale $r_1$ with evaluated quantities on the lattice, 
 we use the values of $r_1/a$ given in Table VIII of Ref.~\cite{Bazavov:2014pvz}.
These values are obtained from the interpolation of the calculated $r_1/a$ values \cite{Bazavov:2014pvz}.
Since $r_1/a$ is a smooth function of the gauge coupling, the errors in the determination
of $r_1/a$ can be largely reduced by using a smooth interpolation (see Ref.~\cite{Bazavov:2014pvz}
for details). We also performed the analysis using the calculated value of $r_1/a$ at each
$\beta$ and checked that our final result does not change.

Now, we can determine the bare charm quark mass $m_{c0}$ on the lattice and the mass ratio of the charm to strange quarks.
To determine $d_0$ and $b$ at each $\beta$, we calculate the masses of $J/\psi$ and $\eta_c$ 
mesons for several trial values of the charm quark mass in the range that 
encompasses the physical value of the charm quark mass. Some details on the determination of the
charmonium masses are given in the Appendix. We perform interpolation of the calculated spin averaged mass
$\overline{M}$ to its physical value.
We find that the linear fits $a\overline{M}=d_0+b(am_{ct})$ work very well, and the statistical errors of $d_0$ and $b$ 
are estimated using the bootstrap analysis.
The results of $m_{c0}$ are also shown in Table \ref{tab:param_charm} with the number of trajectories used 
to calculate $J/\psi$ and $\eta_c$ masses.  The statistical errors 
on $m_{c0}$ are estimated from the errors on $d_0$ and $b$, as well as from the errors on $r_1/a$ added in quadrature.
Table \ref{tab:param_charm} shows that the determination of $m_{c0}$ achieves an accuracy of 0.4\%, 
except for the highest $\beta$ value, where the error becomes about 0.7\% mainly due to the uncertainty of $r_1/a$.
The determined $m_{c0}$ values are used for the calculations of the moments of the pseudoscalar charmonium correlator 
discussed below.

\begin{figure}[tb]
\begin{center}
\includegraphics[width=80mm]{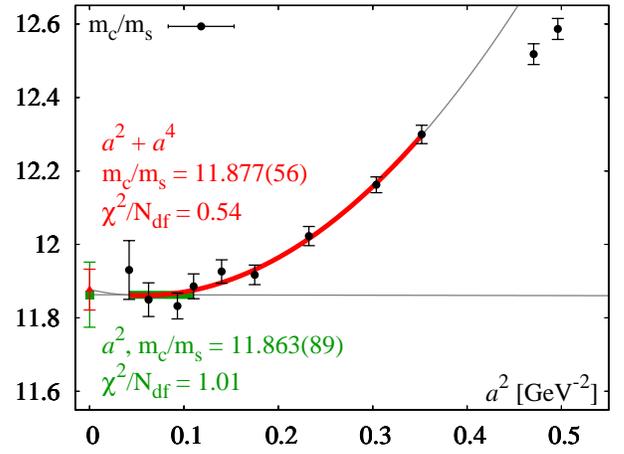}
\caption{The lattice spacing dependence of $m_c/m_s$ together
with continuum extrapolations. The triangle (square) corresponds to the $a^2$
($a^2+a^4$) continuum extrapolation.
The results of extrapolations with $\chi^2/N_{\rm df}$ of the fits are also shown.
The thick lines show the extrapolation curves in the interval in which the fits
have been performed, while the thin lines show the same curves outside that interval.
}
\label{fig:mcms}
\end{center}
\end{figure}

Using the value of $B_0$ calculated from the values of $a m_s$ and 
$a M_{\eta_{s\bar{s}}}$ given in Tables III and V of Ref.~\cite{Bazavov:2014pvz}, $B_0=(a M_{\eta_{s\bar{s}}})^2/am_s$, and
we can determine $m_c/m_s$ using Eq.~(\ref{basic}) at each lattice spacing. 
The error on $B_0$ comes from the error on the mass of the unmixed 
pseudoscalar $s\bar{s}$ meson determined in Ref.~\cite{Bazavov:2014pvz}. This error was included
in the analysis, however, it is subdominant.
The results are shown in Fig.~\ref{fig:mcms} as a function of $a^2$.
At this point, we do not include the errors on $\overline M$, $M_{\eta_{s\bar{s}}}$, and the physical
value of $r_1$, as these are common for all data points.
We performed continuum extrapolations using the $a^2$ form as well as the $a^2$ form plus $a^4$ term. 
These are shown in Fig.~\ref{fig:mcms}. 
The coarsest two lattice spacings are not in the scaling regime and therefore
are not included in the final analysis. Using $a^2+a^4$ extrapolation we obtain
$m_c/m_s=11.877(56)$ with $\chi^2/N_{\rm df}=0.54$, 
while for the $a^2$ extrapolation we have $11.863(89)$ with $\chi^2/N_{\rm df} = 1.01$. 
Since the two extrapolations agree within
the errors, we take $m_c/m_s=11.877(56)$ as our final continuum result. 
Now, adding the errors from the absolute values of $\overline M$, $M_{\eta_{s\bar{s}}}$,
and $r_1$ we obtain our final result:
\begin{equation}
\frac{m_c}{m_s}=11.877 \, (56) \, (72) \,,
\end{equation}
where the first (second) parenthesis indicates the statistical (scale) uncertainty.

\begin{figure}[tb]
\begin{center}
\includegraphics[width=80mm]{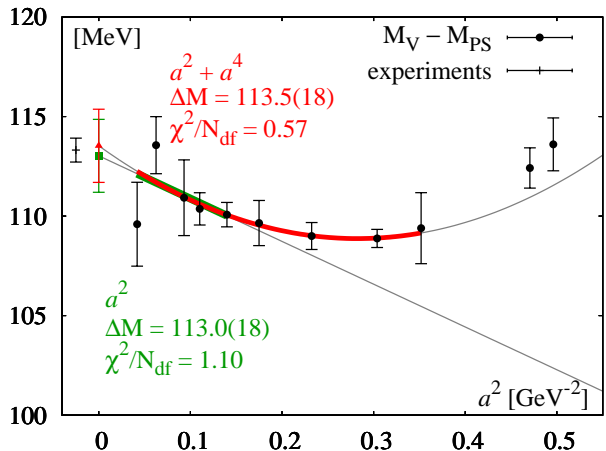}
\caption{The lattice spacing dependence of the charmonium hyperfine splitting together
with continuum extrapolations. The triangle (square) plot corresponds to the $a^2$
($a^2+a^4$) continuum extrapolation.
The results of extrapolations with $\chi^2/N_{\rm df}$ of the fits as well as the experimental value $M_{J/\psi}-M_{\eta_c} = 113.3(6)$ MeV are also shown.
The thick lines show the extrapolation curves in the interval in which the fits
have been performed, while the thin lines show the same curves outside that interval.
}
\label{fig:hyperfine}
\end{center}
\end{figure}

To test possible cutoff effects in our calculations, we consider the charmonium spectra and hyperfine splitting in the continuum limit.
For the pseudoscalar (PS) and vector (V) masses with the trial $am_c$ values, we perform similar fits with Eq.~(\ref{eq:Mbar})
and obtain $d_0^{(i)}$ and $b^{(i)}$ with $i=$PS and V.
Then, those masses on the lattice can be determined as
\begin{equation}
M_i = r_1^{-1} \left(\frac{r_1}{a}\right)d_0^{(i)} + b^{(i)}m_{c0}, \ \ i=\text{PS and V}.
\end{equation}
We find that the dependence of the masses on the lattice spacing becomes very mild. 
From the continuum extrapolation using the $a^2$ form and the six highest $\beta$ values, we obtain
\begin{eqnarray}
M_{\rm PS} &=& 2.982 \, (12) \ {\rm GeV} \, ,\\ 
M_{\rm V}  &=& 3.095 \, (12) \ {\rm GeV} \, ,
\end{eqnarray}
which agree well with the experimental values 
$M_{\eta_c} = 2.9836(6)$ GeV and $M_{J/\psi} = 3.096916(11)$ GeV.
The hyperfine splitting $\Delta M \equiv M_{\rm V}-M_{\rm PS}$ can 
also be extracted from the lattice masses and is shown in Fig.~\ref{fig:hyperfine}.
We perform continuum extrapolations of the hyperfine splitting using $a^2$ and $a^2+a^4$ forms, which are also shown in Fig.~\ref{fig:hyperfine}.
The results from the two continuum extrapolations agree very well within the errors. 
Taking the results from the $a^2+a^4$ extrapolations 
and considering the errors from the absolute values of $\overline{M}$ and $r_1$, we obtain:
\begin{equation}
\Delta M = 113.5 \, (18) \, (7) \ {\rm MeV}.
\end{equation}
This agrees very well with the experimental value $M_{J/\psi}-M_{\eta_c} = 113.3(6)$ MeV.

\begin{figure}[tb]
\begin{center}
\includegraphics[width=80mm]{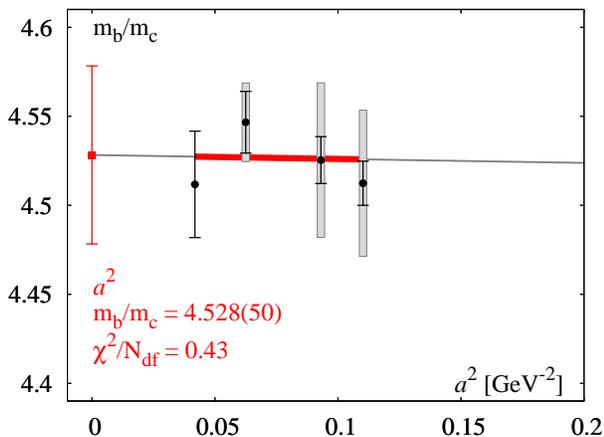}
\caption{The lattice spacing dependence of $m_b/m_c$.
The error bars indicate the statistical errors, 
whereas the gray shadows indicate systematic uncertainties due to the heavy quark mass extrapolations (see text for more details).
Result of the continuum extrapolation done by the $a^2$ form is also shown with $\chi^2/N_{\rm df}$ value of the fit.
The thick line shows the extrapolation curve in the interval in which the fit
has been performed, while the thin line shows the same curve outside that interval.
}
\label{fig:mbmc}
\end{center}
\end{figure}

In the end of the subsection, we determine the ratio of the bottom to charm quark mass $m_b/m_c$.
For this purpose, we chose heavier valence masses than the charm quark mass,
$m_h > m_{c0}$, and calculate the pseudoscalar meson mass $M_{\eta_h}$.
For the four finest lattice spacings at $\beta=7.280$--7.825, we calculate the meson correlation functions at $0.5\le am_h\le1.0$
and determine the quark mass dependence of $M_{\eta_h}$ according to Eq.~(\ref{eq:metah}).
The number of trajectories we use for this calculation is summarized in Table \ref{tab:param_bottom} for each gauge coupling $\beta$.
The details of the extraction of the heavy pseudoscalar meson masses are discussed 
in the Appendix.
The parameters $d_{h0}$ and $b_h$ are determined by the linear fittings of $M_{\eta_h}$ 
as a function of $m_h$, with three data at $am_h=0.7$, 0.8 and 0.9.
In our calculations, 
the interpolation from $M_{\eta_h}$ to the experimental $M_{\eta_b}$ is possible for $\beta=7.825$.
For other $\beta$ values, however, the bottom quark mass corresponds to the region $a m_h>1$, and extrapolations
are necessary to obtain $M_{\eta_b}$.
We estimate uncertainties coming from the extrapolation to the bottom quark masses in the following way: 
First, by using $d_{h0}$ and $b_h$ obtained from the fit 
at $am_h=(0.7,0.8,0.9)$, we determine the bare bottom quark mass $m_{b0}$ at each $\beta$ as
\begin{equation}
m_{b0} = \frac{1}{b_h} \left[ M_{\eta_b} - d_{h0} \left(\frac{r_1}{a}\right) r_1^{-1} \right]
.
\end{equation}
Then, we iterate the fit with the different data points at $0.5\le am_h\le 1.0$, e.g., $am_h=(0.8, 0.9, 1.0)$, and calculate $m_{b0}$ again.
The difference between these two values provides an estimate of the systematic errors.
The numerical values of $m_b/m_c$ obtained from Eq.~(\ref{eq:mbmc}) are summarized 
in Table \ref{tab:param_bottom}, where the first  and second parentheses indicate the statistical and systematic errors, respectively.
The lattice spacing dependence of $m_b/m_c$ is shown in Fig.~\ref{fig:mbmc}, 
where the error bars and gray shadows indicate the statistical and systematic errors, respectively.
We find that the lattice spacing dependence is very mild. 
The systematic errors become larger on coarser lattices and are significantly larger than the statistical errors for $\beta \le 7.373$.
Although there is no significant lattice spacing dependence, we perform the continuum extrapolations with the $a^2$ form,
including the uncertainties from the statistical and systematic errors in quadrature.
As a consequence we obtain
\begin{equation}
\frac{m_b}{m_c} = 4.528 \, (50) \, (27) \, ,
\end{equation}
where the number in the first parenthesis shows the combined statistical and extrapolation errors, and the number in the 
second parenthesis is the combined error from the values of $\overline{M}$, $M_{\eta_b}$ and $r_1$.
As a test of our approach, we have performed the calculations of $m_b/m_c$ by using vector bottomonium $M_V$ masses obtained on the 
lattice combined with the experimental $\Upsilon$ mass and obtained the value $m_b/m_c=4.531(52)$, which is essentially the same as
 above.
The small lattice spacing dependence of the ratio $m_b/m_c$ may appear somewhat surprising. 
Note, however, that the discretization errors for the HISQ action are small also in the heavy quark mass region 
as long as $a m_h \le 1$ \cite{Follana:2006rc}.
In particular, the cutoff dependence of the ground state quarkonium masses was studied in a wide range
of quark masses and was found to be small \cite{McNeile:2010ji}. This is the reason why the 
cutoff dependence of $m_b/m_c$ is small.

\begin{table}[tb]
 \begin{center}
 \caption{The gauge coupling ($\beta$) and the number of trajectories (traj.) used to calculate the bottomonium correlation functions 
 with the corner-wall sources. Also shown are the results of the ratios of the bottom to charm quark mass.  
 The first parenthesis in the last column indicates the statistical errors, 
 and the second one indicates the systematic errors due to extrapolations to heavy quark masses (see the text for more detail).
 }
 \label{tab:param_bottom}
 {\renewcommand{\arraystretch}{1} \tabcolsep = 1mm
 \newcolumntype{.}{D{.}{.}{2}}
 \begin{tabular}{ccl}
 \hline\hline
 \multicolumn{1}{c}{$\beta$} & 
 \multicolumn{1}{c}{traj.}&
 \multicolumn{1}{c}{$m_b/m_c$}\\
 \hline
7.280 & 1800 & 4.512(12)(41)  \\ 
7.373 & 1800 & 4.525(13)(43)  \\ 
7.596 & 1680 & 4.547(17)(22)  \\ 
7.825 & 2562 & 4.512(30)(0)   \\ 
 \hline
 \end{tabular}}
 \end{center}
\end{table}

\subsection{Strong coupling constant and quark masses from the moments}

The strong coupling constant is determined from the moments of the pseudoscalar charmonium correlators on the lattice 
combined with the perturbative expansion of the corresponding quantity.
We calculate the pseudoscalar charmonium meson correlators 
with the valence mass $m_{c0}$ determined above using point sources with large statistics.
Gauge couplings $\beta$ and the number of trajectories used to calculate the moments 
are summarized in Table \ref{tab:mom} (see also Table \ref{tab:param_charm} for the corresponding lattice size and $m_{c0}$).
The numerical results of the reduced moments are also summarized in Table \ref{tab:mom} up to $n\le10$.
The statistical errors of the reduced moments $R_n$ have been estimated using jackknife procedure, and
we checked again that there is no dependence on the jackknife bin size.

\begin{figure}[tb]
\begin{center}
\includegraphics[width=80mm]{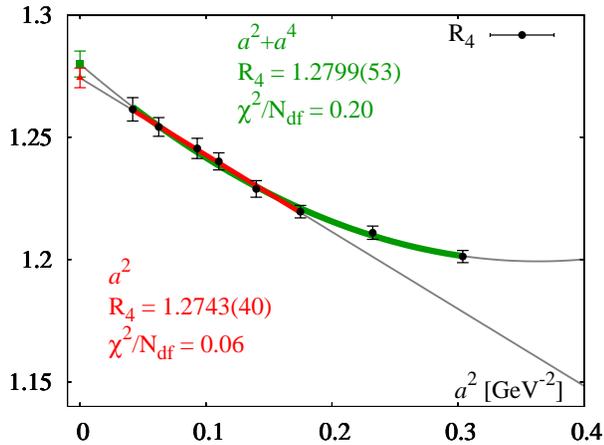} 
\end{center}
\caption{Lattice results for $R_4$. Also shown are the continuum extrapolations.
The thick lines show the extrapolation curves in the interval in which the fits
have been performed, while the thin lines show the same curves outside that interval.
}
\label{fig:r4}
\end{figure}

Since the lattice gauge configurations used in our study do
not include the effects of charm quarks, following Ref.~\cite{Allison:2008xk}
we estimate such effects using perturbation theory. It was shown 
that charm quarks increase the value of $R_4$ by $0.7\%$ \cite{Allison:2008xk}. 
Therefore,
we scale our lattice results for $R_4$ by $1.007$.
In the following, we always give the rescaled value of $R_4$.
The results of $R_4$ are shown in Fig.~\ref{fig:r4} as a function of the lattice spacing $a^2$.
The lattice spacing dependence of $R_4$ is significant, and for
the coarsest lattice, it amounts to $6\%$.
We have performed continuum extrapolation of our results using various fit forms.
For HISQ action, the leading discretization effects are expected to be 
$\alpha_s a^2$ and $a^4$. It is usually assumed that the running of the 
coupling constant $\alpha_s$ can be neglected if the considered range of the lattice
spacing is not too large. Therefore, we can fit the numerical results for
$R_4$ using $a^2$ and $a^2+a^4$ forms. We could also perform continuum extrapolations
using the $\alpha_s a^2$ form by defining the boosted coupling constant
\begin{equation}
\alpha_s^b(1/a) = \frac{1}{4\pi}\frac{g_0^2}{u_0^4} \, ,
\end{equation}
where $g_0^2=10/\beta$ is a bare lattice gauge coupling and $u_0$ is an averaged link valuable defined by the plaquette $u_0^4 = \langle {\rm Tr}U_\square\rangle/3$.
Furthermore, we perform continuum extrapolations using $\alpha_s a^2+ a^4$ and $\alpha_s (a^2+ a^4)$ forms . 
We obtain the following continuum results:
\begin{equation}
\begin{array}{cl}
R_4 = 1.2743(40) \, , & a^2 \ {\rm fit}\\
R_4 = 1.2799(53) \, , & a^2 + a^4\ {\rm fit}\\
R_4 = 1.2705(37) \, , & \alpha_s^b a^2 \ {\rm fit}\\
R_4 = 1.2769(49) \, , & \alpha_s^b a^2 + a^4\ {\rm fit}\\
R_4 = 1.2759(47) \, , & \alpha_s^b (a^2 + a^4 )\ {\rm fit}
\end{array}
\end{equation}
The continuum extrapolations with $a^2$ and $a^2+a^4$ forms are shown in Fig. \ref{fig:r4}.
When performing fits with $a^2$ and $\alpha_s^b a^2$ forms, the data 
for the two coarsest lattice spacings have been excluded since these are not in the scaling regime.
As our final continuum result, we take the value of $R_4$ obtained from the simple $a^2$ extrapolation
and assign a systematic  error of $0.0047$ due to the continuum extrapolation to take into account the
spread in the central values of $R_4$ obtained above:
\begin{equation}
R_4=1.2743(40)(47).
\end{equation}
Our continuum result for $R_4$ agrees with the continuum results
$R_4=1.281(5)$ and $R_4=1.282(4)$ obtained in Ref.~\cite{Allison:2008xk} and Ref.~\cite{McNeile:2010ji},
respectively, within errors. Our central value is slightly smaller. 
It should be pointed out that we have many more lattice spacings in the region $a<0.1$ fm to perform
the continuum extrapolations compared to Refs. \cite{Allison:2008xk} and \cite{McNeile:2010ji}.

Using the above results for $R_4$ as well as the corresponding perturbative expansion, it is
straightforward to determine $\alpha_s(\mu=m_c)$. We obtain the value
\begin{equation}
\alpha_s(\mu=m_c, n_f=3)=0.3697 \, (54) \,(64) \, (15) \, ,
\label{eq:alphas_mc}
\end{equation}
where the first error is statistical,  the second error corresponds to the uncertainty of the continuum extrapolation,
and the last error comes from the truncation of
the perturbative series for $r_4$. The truncation error was estimated as follows: First, we used
the perturbative result up to order $\alpha_s^3$ to estimate the strong coupling constant.
Then we included $\alpha_s^4$ term with a coefficient equal to $r_{43} \times 2$ and estimated
the strong coupling constant again. The difference between these two estimates is the truncation error.

\begin{figure}[tb]
\begin{center}
\includegraphics[width=80mm]{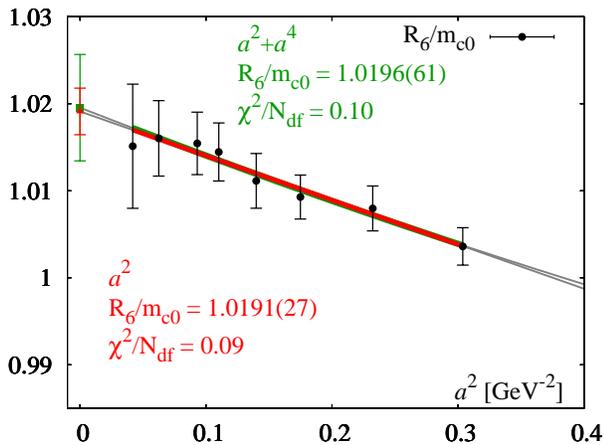}
\end{center}
\caption{Lattice results for $R_6$. Also shown are the continuum extrapolations.
The thick lines show the extrapolation curves in the interval in which the fits
have been performed, while the thin lines show the same curves outside that interval.
}
\label{fig:r6}
\end{figure}

Now let us discuss the determination of the charm quark mass. The value of the charm
quark mass is also needed to specify $\alpha_s$ at any scale using Eq.~(\ref{eq:alphas_mc}). 
We determine the charm quark mass by considering $R_6$. 
The $\overline{\rm MS}$ charm mass can be estimated from the lattice mass as
$m_c=r_6/(R_6/m_{c0})$.
The effect of charm quark loops turns out to be much smaller for $R_6$ than for $R_4$
when estimated using perturbation theory. It was estimated that charm quark loops
increase $R_6$ and this effect amounts to
$0.1\%$ \cite{Allison:2008xk}. 
We corrected our lattice data for this tiny effect and always show the corrected values of $R_6$ in
the discussions below.
Our lattice results for $R_6/m_{c0}$ are shown in Fig. \ref{fig:r6}. 
The lattice spacing dependence of $R_6/m_{c0}$ is milder than for $R_4$. The maximal discretization errors
are less than $2\%$. This is not surprising since 
the contribution to $G_6$ from 
the data points at small $t$ is smaller than for $G_4$.
The data points at small $t$ 
are the most sensitive to the lattice artifacts.

To obtain the continuum result, we again perform $a^2$ extrapolation
which results in 
$R_6/m_{c0}=1.0191(27)$.
Performing an extrapolation with the $a^4$ term included results in $R_6/m_{c0}=1.0196(61)$,
which is in very
good agreement with the above result (shown in Fig.~\ref{fig:r6}). We also performed extrapolations using
$\alpha_s a^2$ and $\alpha_s a^2+a^4$ forms and obtained $R_6/m_{c0}=1.0181(23)$ and
$R_6/m_{c0}=1.0192(56)$, respectively. Since all the above continuum results agree well
within the estimated statistical errors, we use the value from simple $a^2$ extrapolations
as our final continuum estimate
\begin{equation}
R_6/m_{c0}=1.0191(27).
\end{equation}
Using this and the 
perturbative result for $r_6$, we obtain
\begin{equation}
m_c(\mu=m_c, n_f=3)=1.2668(33)(34)(79)(73) \ {\rm GeV},
\end{equation}
where the first error is statistical, the second error is the truncation error in $r_6$, the third
error comes from $\alpha_s$ determined above, and the last error comes from setting the scale in
our lattice calculations.

The higher moments can also be utilized to determine $\alpha_s$  and $m_c$. Namely, we can use 
$R_{n-2}/R_n$ to determine $\alpha_s$ and  $R_n/m_{c0}$  to determine $m_c(m_c)$. Here, $n\ge8$.
These calculations provide a valuable cross check  for the extraction of $\alpha_s$ and $m_c$.
There is also an advantage that the lattice spacing dependence for higher moments is expected to become milder as discussed above. 
Thus, more accurate continuum extrapolations could be possible.
There is, however, a disadvantage in using higher moments. Higher order contributions in the perturbative expansion become significant
for higher moments, but the perturbative coefficients are known up to $\alpha_s^3$ orders at present.
The absence of higher order perturbative calculations leads to larger truncation errors.
On the lattice side there is also a disadvantage that the higher moments require information 
of the correlation functions at larger distance [c.f. Eq.~(\ref{eq:latmom})], but the calculation is performed on the finite lattice.
Thus, the  results for the higher moments potentially suffer from larger finite volume effects. In our calculations,
the finite volume effects become more serious at finer lattice spacing.

In our study, the finite volume effects mostly appear in the moments of the free correlation functions $G_n^{(0)}$ introduced in Eq.~(\ref{eq:redmom}).
This is due to the fact that the exponential decay of the free meson correlator is governed 
by $2 m_{c0} \simeq 1.8-2.0$ GeV rather than by $m_{\eta_c} \simeq 3$ GeV.
To investigate such effects, we calculate the free moments in the infinite temporal-size 
limit by using the results of $1.5N_t$ and $2N_t$ and estimate the finite 
volume effects from $G_n^{(0)}(N_t)/G_n^{(0)}(\infty)$, 
where the numerator is the same one used in Eq.~(\ref{eq:redmom}), whereas the denominator is that on the infinite temporal size.
We perform this calculation for the highest two $\beta$ values 
and find that the effects become negligible for $n\le6$, 
whereas 1\% and 6\% effects appear for $R_8$ and $R_{10}$ at $\beta=7.596$, respectively,
and 7\% and 23\% effects appear for $R_8$ and $R_{10}$ at $\beta=7.825$, respectively.
Those are larger than the amount of the statistical errors of corresponding quantities and we omit the corresponding data in our analysis.
We perform the continuum extrapolations of $R_{n-2}/R_n$ and $R_n/m_{c0}$ for $n=8$ and 10 using $a^2$ form and including
the lattice data for $\beta=7.030$--7.373 and obtain the following continuum results:
\begin{eqnarray}
R_6/R_8       &=& 1.1140(18),\\
R_8/R_{10}    &=& 1.04954(65),\\
R_8/m_{c0}    &=& 0.9167(54),\\
R_{10}/m_{c0} &=& 0.8731(50).
\end{eqnarray}
Since the continuum extrapolations using the $a^2+a^4$ fit form with the lattice data for $\beta=6.740$--7.373 leads to results
that agree very well with the above results within errors, we consider them as our final continuum results.
From the continuum results for the ratio $R_{n-2}/R_n$ and $R_n/m_{c0}$, we obtain the following
values for $\alpha_s$ and $m_c$:
\begin{equation}
\begin{array}{cccc}
              & &    n=8               &     n=10        \\
\alpha_s(m_c) &=& 0.3954(71)(210)   \, & 0.3611(50)(152) \\
m_c(m_c)      &=& 1.2717(75)(9)     \, & 1.2708(73)(35) \, ,
\end{array}
\end{equation}
where the first (second) parenthesis indicates the statistical (truncation) errors.
We see that the truncation errors for $\alpha_s$ are an order of magnitude larger than
the truncation errors coming from $R_4$. Within the large errors, the above values
of $\alpha_s$ are consistent with the $\alpha_s$ determination from the fourth moment. On
the other hand, the truncation errors are fairly small for the charm quark mass, and the above
values of $m_c$ agree well with our previous determination.

\begin{table}[tb]
 \begin{center}
 \caption{The gauge couplings ($\beta$) and the number of trajectories (traj.) used for the calculations 
  of the moments of the pseudoscalar correlation functions with the valence mass $m_{c0}$ and point sources.
 Numerical results of the reduced moments $R_n$ are also shown up to $n\le10$.}
 \label{tab:mom}
 {\renewcommand{\arraystretch}{1} \tabcolsep = 1mm
 \newcolumntype{.}{D{.}{.}{2}}
 \begin{tabular}{ccllllll}
 \hline\hline
 \multicolumn{1}{c}{$\beta$} & 
 \multicolumn{1}{c}{traj.}&
 \multicolumn{1}{c}{$R_4$}&
 \multicolumn{1}{c}{$R_6$}&
 \multicolumn{1}{c}{$R_8$}&
 \multicolumn{1}{c}{$R_{10}$}&
 \\
 \hline
6.740  & 8005  & 1.2012(24)   & 1.0252(12)   & 0.94261(67)  & 0.89873(48) \\ 
6.880  & 8095  & 1.2110(27)   & 1.0014(13)   & 0.91399(75)  & 0.87025(54) \\ 
7.030  & 9830  & 1.2196(25)   & 0.9763(12)   & 0.88726(69)  & 0.84523(50) \\ 
7.150  & 7902  & 1.2289(33)   & 0.9577(16)   & 0.86794(83)  & 0.82661(58) \\ 
7.280  & 8058  & 1.2401(34)   & 0.9424(17)   & 0.85182(90)  & 0.81119(63) \\ 
7.373  & 9246  & 1.2454(41)   & 0.9303(18)   & 0.84022(94)  & 0.80054(65) \\ 
 \multicolumn{1}{l}{7.596} & 
 \multicolumn{1}{l}{9510}&
 \multicolumn{1}{l}{1.2542(38)}&
 \multicolumn{1}{l}{0.9020(17)}&
 \multicolumn{1}{c}{--}&
 \multicolumn{1}{c}{--} \\
 \multicolumn{1}{l}{7.825} & 
 \multicolumn{1}{l}{9516}&
 \multicolumn{1}{l}{1.2614(47)}&
 \multicolumn{1}{l}{0.8811(20)}&
 \multicolumn{1}{c}{--}&
 \multicolumn{1}{c}{--} \\
\hline
 \end{tabular}}
 \end{center}
\end{table}
Before concluding this section, let us compare the continuum results for the higher moments $R_n$, $n\ge 6$ with
the HPQCD results \cite{Allison:2008xk,McNeile:2010ji}. 
In Refs. \cite{Allison:2008xk,McNeile:2010ji} a slightly different definition of $R_n$ was used: 
 $R_n=m_{\eta_c}/(2 m_{c0}) (G_n/G_n^{(0)})^{1/(n-4)}$.
If we use this definition, we obtain $R_6=1.520(4)$, $R_8=1.367(8)$, and $R_{10}=1.302(8)$, which agree well  with the
results of Ref.~\cite{McNeile:2010ji}:
$R_6=1.527(4)$, $R_8=1.373(3)$, and $R_{10}=1.304(2)$, as well as with the results of Ref.~\cite{Allison:2008xk}:
$R_6=1.528(11)$, $R_8=1.370(10)$, and $R_{10}=1.304(9)$.

\section{Summary of results and comparison with other works}
\label{sec:comp}

Now, we summarize the main findings of this paper.
From the masses of pseudoscalar and vector mesons on the lattice we obtained the quark mass ratios:
\begin{equation}
\frac{m_c}{m_s}=11.877(91) \, , \ \ \ \
\frac{m_b}{m_c}=4.528(57) \, , \label{eq:ratio}
\end{equation}
where the statistical and systematic errors are added in quadrature.
On the other hand, from the moments of the pseudoscalar charmonium correlation functions, we estimated 
 the strong coupling constant and the charm quark masses in $\overline{\rm MS}$ scheme for $\mu=m_c$:
\begin{eqnarray}
\alpha_s(\mu=m_c, \, n_f=3) &=& 0.3697 (85) \, , \label{eq:als} \\
     m_c(\mu=m_c, \, n_f=3) &=& 1.267 (12) \ {\rm GeV} \,.  \label{eq:mc}
\end{eqnarray}

\begin{figure}[tb]
\begin{center}
\includegraphics[width=60mm]{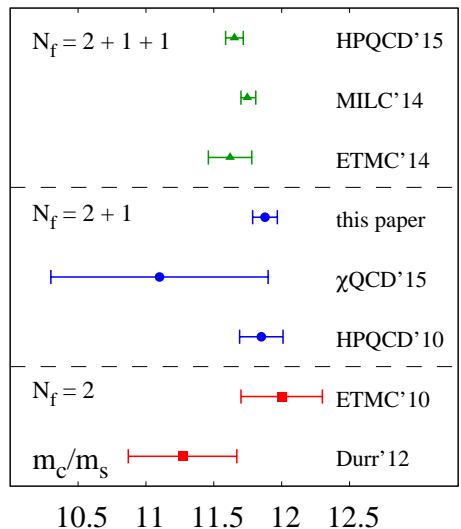}
\end{center}
\caption{Determinations of the ratio of the charm quark mass to strange quark mass $m_c/m_s$ in lattice QCD simulations.
We include the determinations from
HPQCD'15 \cite{Chakraborty:2014aca}, 
MILC'14  \cite{Bazavov:2014wgs},
ETMC'14  \cite{Carrasco:2014cwa},
$\chi$QCD'15 \cite{Yang:2014sea},
HPQCD'10 \cite{Davies:2009ih},
ETMC'10  \cite{Blossier:2010cr}, and
Durr'12  \cite{Durr:2011ed}.
}
\label{fig:mcms_comp}
\end{figure}

\begin{figure}[tb]
\begin{center}
\includegraphics[width=60mm]{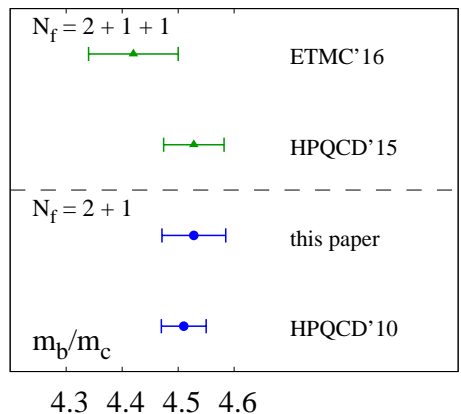}
\end{center}
\caption{Determinations of the ratio of the bottom quark mass to charm quark mass $m_b/m_c$ in lattice QCD simulations.
In the comparison we include the results from 
ETMC'16  \cite{Bussone:2016iua},
HPQCD'15 \cite{Chakraborty:2014aca}, and
HPQCD'10 \cite{McNeile:2010ji}.}
\label{fig:mbmc_comp}
\end{figure}

Let us first compare the quark mass ratios, which are scale and scheme independent quantities,  with other lattice determinations.
In Fig.~\ref{fig:mcms_comp}, we show our result on $m_c/m_s$  and compare it with several recent lattice QCD results.
We find that the results in 2+1 flavor simulations show similar values.  Our result for $m_c/m_s$ is about one sigma larger than 2+1+1 flavor results.
In Fig.~\ref{fig:mbmc_comp}, we compare our result on $m_b/m_c$ with other recent lattice QCD determinations.
Our result agrees well with the result from the HPQCD Collaboration and has similar errors.
By just multiplying both ratios, we obtain the mass ratio of the bottom to the strange quarks:
\begin{equation}
\frac{m_b}{m_s} = 53.78(79) \, .
\end{equation}
This can be compared with one of the prediction in the grand unified theory, namely, the Georgi-Jarlskog relation which
states: $m_b/m_s = 3m_\tau/m_\mu = 50.45$ \cite{Georgi:1979df}.
Our result is 6\% away from this prediction.

\begin{figure}[tb]
\begin{center}
\includegraphics[width=60mm]{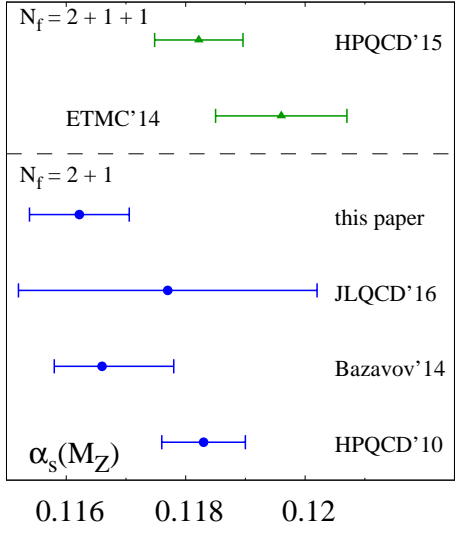}
\end{center}
\caption{Determinations of $\alpha_s(M_Z,n_f=5)$ in lattice QCD simulations.
Recent results are included from 
HPQCD'15     \cite{Chakraborty:2014aca}, 
ETMC'14      \cite{Blossier:2013ioa},
JLQCD'16     \cite{Nakayama:2016atf},
Bazavov'14   \cite{Bazavov:2014soa}, and
HPQCD'10     \cite{Davies:2009ih}.
}
\label{fig:alsmz_comp}
\end{figure}

To compare our 
result on the strong coupling constant with other determinations, we need to evolve it to higher scales $\mu$.
We do so by using the 4-loop perturbation theory in the $\overline{\rm MS}$ scheme and the RunDeC package \cite{Chetyrkin:2000yt}.
First, we compare our result to two low energy determinations of $\alpha_s$ in 2+1 flavor lattice QCD simulations.
One of them comes from the analysis of static quark-antiquark energy \cite{Bazavov:2012ka,Bazavov:2014soa} with
the most recent value $\alpha_s(1.5 {\rm GeV})=0.336(+12)(-0.008)$ \cite{Bazavov:2014soa}.
Evolving our result to $\mu=1.5$ GeV and propagating the uncertainties, we obtain
\begin{equation}
\alpha_s(1.5 \, {\rm GeV}, n_f=3)=0.3316(69)
\end{equation}
in excellent agreement with the above result.
Another low energy determination of $\alpha_s$ by the HPQCD Collaboration comes from the 
lattice calculations of the moments of meson correlators consisting of heavy quarks with several
values of the heavy quark mass and the Bayesian fit of these correlators to the perturbative result 
\cite{McNeile:2010ji}: $\alpha_s(5 {\rm GeV}, n_f=3)=0.2034(21)$. Evolving our results to $\mu=5 $ GeV
and propagating the errors, we obtain 
\begin{equation}
\alpha_s(5 \, {\rm GeV}, n_f=3)=0.1978(22) ,
\end{equation} 
which is two sigma lower than the above result.
Furthermore, we evolve our result to the commonly used scale $\mu=M_Z$ 
with $n_f=5$ by adding the contributions of the charm and bottom quarks
using the RunDeC package\footnote{
In more detail, we start from our $\alpha_s(m_c,n_f=3)$ and 
match it to $\alpha_s(m_c, n_f=4)$ using $m_c(m_c)$ as the threshold.
We evolve the corresponding $\alpha_s(m_c,n_f=4)$ to the scale $\mu=m_b(m_b)$, where we match
it to $\alpha_s(m_b,n_f=5)$. Finally, we evolve the coupling constant to $\mu = M_Z$.}
\begin{equation}
\alpha_s (M_Z , n_f=5) = 0.11622(84) \, .
\label{eq:alsmz}
\end{equation}
In Fig.~\ref{fig:alsmz_comp}, we compare our $\alpha_s(M_Z)$ with recent results obtained by other collaborations.
We see that our result is in  agreement with Ref.~\cite{Bazavov:2014soa} (``Bazavov'14'' )  but is lower
than other lattice determinations.

\begin{figure}[tb]
\begin{center}
\includegraphics[width=60mm]{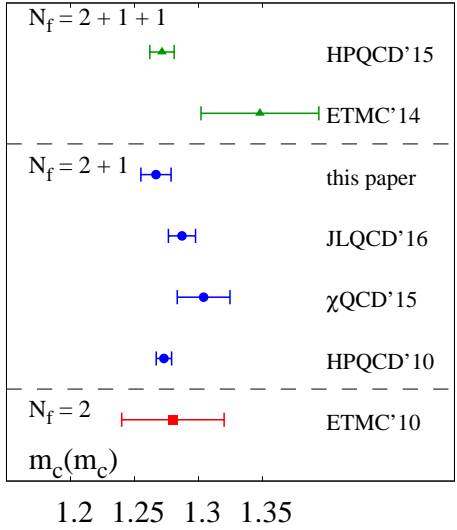}
\end{center}
\caption{Charm quark mass in $\overline{\rm MS}$ scheme $m_c(\mu=m_c)$ obtained
in this work and compared with other lattice determinations:
HPQCD'15 \cite{Chakraborty:2014aca}, 
ETMC'14  \cite{Carrasco:2014cwa},
$\chi$QCD'15 \cite{Yang:2014sea}
HPQCD'10 \cite{Davies:2009ih}, and
ETMC'10  \cite{Blossier:2010cr}.
}
\label{fig:mcmc_comp}
\end{figure}

The charm quark mass $m_c(m_c)$ was directly obtained by extrapolating the lattice moments to the continuum and matching 
those to the corresponding perturbative results.
Figure \ref{fig:mcmc_comp} shows the comparison of our charm quark mass with the recent results of other collaborations.
We find that our result is similar to other lattice determinations but is lower than the ETMC '14 result \cite{Carrasco:2014cwa}.
We can also calculate the strange and bottom quark masses from $m_c(m_c)$ with the quark mass ratios and obtain
\begin{eqnarray}
m_s(\mu=2 \, {\rm GeV}, \, n_f=3) &=& 92.0(1.7) \ {\rm MeV} \, , \\
m_b(\mu=m_b, \, n_f=5)            &=& 4.184(89) \ {\rm GeV} .
\label{eq:mb}
\end{eqnarray}
Our result of $m_b(m_b)$ is almost the same as the recent result by the HPQCD Collaboration $m_b(m_b)=4.162(48)$ \cite{Chakraborty:2014aca} and 
also agrees with the ETMC result $m_b(m_b)=4.26(10)$ \cite{Bussone:2016iua} within the errors.
On the other hand, for the strange quark mass, our result agrees well
with the recent HPQCD result \cite{Chakraborty:2014aca} within uncertainties: $m_s(2{\rm GeV})=93.6(8)$ MeV.
It is lower than the values obtained by the ETMC Collaboration \cite{Carrasco:2014cwa}, $m_s(2{\rm GeV})=99.6(4.3)$ MeV and D\"urr {\it et al} \cite{Durr:2011ed},
$m_s(2{\rm GeV})=97.0(2.6)(2.5)$ MeV. Evolving our result to $\mu=3$ GeV, we get $m_s(3{\rm GeV})=83.6(1.5)$. This is in good agreement with 
the result from RBC/UKQCD, $m_s(3{\rm GeV})=81.64(1.17)$ MeV \cite{Blum:2014tka}.

\section{Conclusion}
\label{sec:conclusion}

We have performed determinations of the quark mass ratios as well as the strong coupling constant and the quark masses in 2+1 flavor lattice QCD simulations.
The former have been obtained from the pseudoscalar and vector meson masses together with the experimental mass values,
whereas the latter have been obtained from the moments of the pseudoscalar charmonium correlators and its comparison to the perturbative result
at scale $\mu=m_c$.

At the level of the reduced moments our results agree well with the results obtained by the HPQCD Collaboration.
Our results for bottom, charm and strange quark masses are also in very good agreement with the HPQCD results. We
determined the QCD running coupling constant in the $\overline{\rm MS}$ scheme at the lowest energy scale so far. The error in our determination
of the strong coupling constant is dominated by the lattice error, whereas the error due to the truncation of the perturbative series is very small.
Evolving this low energy determination to $\mu=M_Z$ we obtain $\alpha_s(M_Z,n_f=5)=0.11622(84)$, which is lower than the most lattice QCD
determinations, as well as the PDG value.

One open issue with the present determination based on moments of heavy meson
correlators is whether the systematic error of the perturbative expansion is
sufficiently conservative. Recent analysis seems to suggest that the error in
the perturbative expansion may be underestimated \cite{Dehnadi:2015fra}.

\section*{ACKNOWLEDGMENTS}
This work was supported by U.S. Department of Energy under 
Contract No. DE-SC0012704.
The calculations have been carried out on 
USQCD clusters in Jlab.
We thank Christian Hoebling for useful discussions on the form
of continuum extrapolations.

\appendix
\section{Pseudoscalar meson masses}
\label{app:heavy}
In this appendix, we discuss the determination of the pseudoscalar meson masses
for various quark masses including the quark mass region 
utilized to estimate the bottom quark mass.
To determine the bottom quark mass, we calculate the meson correlator in the pseudoscalar channel 
and estimate the lowest lying pseudoscalar meson masses, $M_{\eta_h}$, at the quark mass range of $0.7 \le am_h \le 0.9$.
Then we extrapolate to the region of heavier quark masses.
Thus, high quality extraction of the ground state meson masses is crucial to ensure the quality of the heavy quark extrapolations.

\begin{figure}[t]
\begin{center}
\includegraphics[width=80mm]{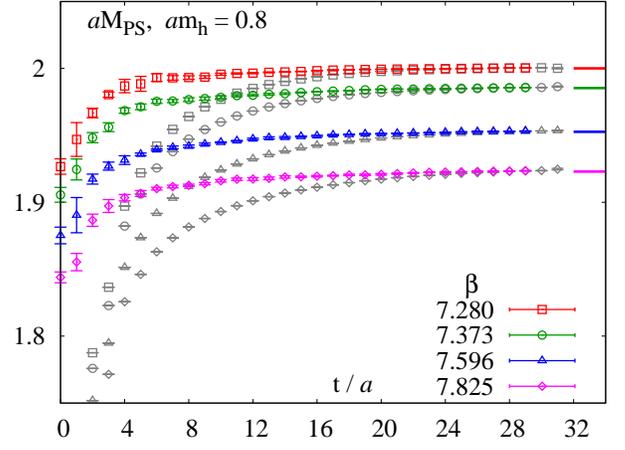}
\end{center}
\caption{Effective masses extracted from the pseudoscalar meson correlators at the valence quark mass of $am_h=0.8$ (gray symbols) 
as a function of the separation $t$. Also shown are the extracted pseudoscalar meson masses as function of $t_{\rm min}/a$. 
The lines in the right-hand-side indicate estimated magnitude of the plateaus for each $\beta$ (see text for more details).}
\label{fig:heavy_efms}
\end{figure}

To demonstrate the quality of meson mass determination in Fig.~\ref{fig:heavy_efms}, we show
the effective masses as well as the fit results for the pseudoscalar channel
at $am_h=0.8$ obtained with corner-wall sources. In the figure, results 
for four $\beta$ values are shown,  which are used to estimate the bottom quark mass in the continuum limit.
Here, the gray symbols correspond to the  results of the effective masses,  whereas 
the colored symbols depict the fit results of the pseudoscalar meson masses performed in the range at [$t_{\rm min}/a$,\, $N_t-t_{\rm min}/a$].
We find that both the effective masses and fit results approach a plateau for large $t$ or large values of $t_{\rm min}/a$.
On finer lattices, one needs larger separations to achieve the plateau. 
For the finest lattice, $\beta=7.825$, the plateau behavior can be seen at $t/a \ge 24$.
The values  of the plateaus 
are estimated by averaging over the fit results  $t_{\rm min}/a=24$--27. This is also shown by the lines in the right-hand side.
We find that the effective masses as well as the fit results are converged to the lines at a large distance which implies that
magnitudes of our extracted plateau well reproduce the lowest lying masses of the corresponding states.

\begin{figure}[t]
\begin{center}
\includegraphics[width=80mm]{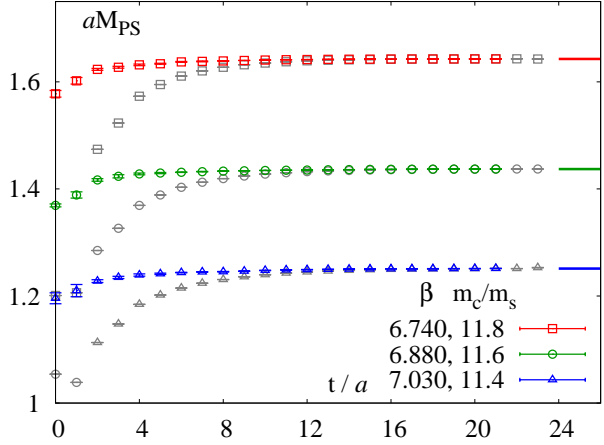}
\end{center}
\caption{Effective masses extracted from the pseudoscalar meson correlators with the valence quark masses around the charm quark mass (gray symbols) 
as a function of the separation $t$. Also shown are the extracted pseudoscalar meson masses as function of $t_{\rm min}/a$. 
The lines in the right-hand side indicate estimated values of the plateaus for each $\beta$ (see text for more details).}
\label{fig:mc_efm}
\end{figure}

A similar analysis has been performed for the valence quark masses around the charm quark mass. The corresponding
results are shown in Fig.~\ref{fig:mc_efm} for several gauge couplings, $\beta$. 
Once again, we see that the extracted masses are stable with respect
to the variation of $t_{\rm min}/a$ within a reasonable range, and the extracted masses agree with the plateaus of the effective
masses.  

\bibliography{pap_bib}
\end{document}